\documentstyle[prl,aps,twocolumn,floats,epsf]{revtex}
\def\geqap{\,\raise 2pt \hbox{$>\kern-11pt \lower 5pt \hbox{$\sim$}$}\,}
\def\leqap{\,\raise 2pt \hbox{$<\kern-10pt \lower 5pt \hbox{$\sim$}$}\,}
\begin{document}
\draft
\twocolumn[\hsize\textwidth\columnwidth\hsize\csname @twocolumnfalse\endcsname
\title{Hole dynamics in spin and orbital ordered vanadium perovskites}
\author{Sumio Ishihara} 
\address{Department of Physics, Tohoku University, Sendai 980-8578, Japan}
\date{\today}
\maketitle
\begin{abstract}
Hole dynamics in spin and orbital ordered vanadates 
with perovskite structure is investigated.  
A mobile hole coupled to the spin excitation (magnon) in the spin G-type and orbital C-type (SG/OC) ordered phase, 
and that to the orbital excitation (orbiton) in the spin C-type and orbital G-type (SC/OG) one 
are formulated on an equal footing. 
The observed fragile character of the (SG/OC) order 
is attributed to the orbiton softening caused by a reduction of the staggered magnetic order parameter. 
It is proposed that the qualitatively 
different hole dynamics in the two spin-orbital 
ordered phases in vanadates can be probed by the optical spectra. 
\end{abstract}
]
\narrowtext
%
%\noindent
%
One of the central issues in the correlated oxides is doped Mott 
insulator \cite{tokura,book}. 
The prototypical compound  
is the high Tc superconducting cuprates (HTSC)
where the parent Mott  
insulating state with the antiferromagnetic (AFM) long-range order is understood by the 
single-band electronic model 
with strong Coulomb interactions. 
Apart from the half filling, 
doped holes are strongly renormalized by interaction with spin excitations. 
%AFM order rapidly collapses 
%and the superconducting phase appears with increasing carrier.  

There is another class of Mott insulator where 
the orbital degeneracy of the transition-metal ion remains. 
Mobile carriers doped into this class of insulator 
couple to the spin and orbital arrays. 
Vanadates with the perovskite structure 
$R_{1-x}$$A_x$VO$_3$ 
($R$: a trivalent rare-earth ion, 
$A$: a divalent alkaline-earth ion) 
are the likeliest candidate. 
A number of researches have been done 
in connection with the spin-orbital orders in 
insulating $R$VO$_3$\cite{ren,miyasaka00,miyasaka03,sawada,khaliullin01,horsch,motome,shen,ulrich,ishihara04}. 
In a V$^{3+}$ ion,  
two electrons occupy the $t_{2g}$ orbitals with the 
$S=1$ high-spin state. 
Two kinds of spin-orbital orders are 
found in the ground state \cite{ren,miyasaka03}: 
the G-type (three-dimensional (3D) staggered) AFM order associated 
with the C-type (rod type) orbital one termed 
the (SG/OC) phase in YVO$_3$, 
and the C-type AFM order with 
the G-type orbital one termed the (SC/OG) phase in LaVO$_3$. 
Both the orders have the commonly occupied $d_{xy}$ orbital at each V site, 
and the alternately occupied $d_{yz}/d_{zx}$ orbitals 
with the AFM spin alignment in the $ab$ plane. 
A difference between the two  
is seen along the $c$ axis; 
the staggered spin (orbital) order associated with the 
uniform orbital (spin) one in the (SG/OC) ((SC/OG)) phase. 

Apart from the well studied $R$VO$_3$, 
little is known about nature of mobile holes 
in the spin-orbital ordered $R_{1-x}A_{x}$VO$_3$.  
A dynamics of a $d_{xy}$ hole 
is only to be expected from the consequence in a well-known doped 
two-dimensional (2D) quantum AFM  
and is common in the (SG/OC) and (SC/OG) phases. 
A clear contrast is anticipated 
in a hole motion along the $c$ axis; 
in the (SG/OC) and (SC/OG) phases, 
a mobile hole is strongly coupled to 
the spin and orbital excitations, 
i.e. magnon and orbital, respectively.  
Orbiton, standing for a quantized object of the collective excitation 
in orbital order, 
is observed in the Raman scattering of LaMnO$_3$ \cite{saitoh} and 
is also proposed in $R$VO$_3$ \cite{ishihara04}. 
Thus, the perovskite vanadates are idea 
for study of the quantum hole dynamics in a spin-orbital 
ordered states.
One of the striking difference between the observed two phases 
is the stability; 
the (SG/OC) phase is rather fragile against 
the hole doping \cite{miyasaka00,miyasaka_un}. 
the (SG/OC) phase realized in YVO$_3$ rapidly 
disappears by doping of holes about 
$x \le 0.05$ in Y$_{1-x}$Ca$_x$VO$_3$. 
This phase is changed into (SC/OG) which is stable, at least, 
up to $x \sim 0.4$. 
The (SC/OG) phase in a series of La$_{1-x}$Sr$_x$VO$_3$ 
also survives until $x=0.178$ \cite{miyasaka03}. 
This fragile nature of (SG/OC)  
is not explained within the doped 3D AFM with $S=1$. 

We present, in this letter, a theory of 
doped perovskite vanadates with spin-orbital orders. 
The hole dynamics associated with magnon and orbiton 
in the (SG/OC) and (SC/OG) phases are formulated 
based on the self-consistent Born approximation (SCBA). 
The different quasi-particle (QP) properties between the two phases 
and the observed barely stable (SG/OC) phase 
are attributed to the softening of the orbital excitation. 

Our starting point is the $t-J$ type model 
with the $t_{2g}$ orbital degree of freedom,  
$ {\cal H}={\cal H}_t+{\cal H}_J+{\cal H}_{JT} $,  
derived from the generalized Hubbard model 
in the large on-site interaction limit \cite{ishihara02}. 
The exchange term 
between nearest neighbor (NN) spins and orbitals has the form 
%\begin{equation}
$
{\cal H}_J={\cal H}_{^4A_2}+{\cal H}_{^2E}+{\cal H}_{^2T_1}+{\cal H}_{^2T_2} 
$
%\label{eq:hj}
%\end{equation}
where 
$^4A_2$, $^2E$, $^2T_{1}$, and $^2T_2$ 
indicate the intermediate $(t_{2g})^3$ configurations of the exchange processes. 
The leading term involves the high-spin $^4A_2$ state explicitly given as   
\begin{eqnarray}
{\cal H}_{^4 A_2}=-\frac{ J_{^4 A_2} }{6}
\sum_{\langle ij \rangle}
\left (  2+ \vec S_i \cdot \vec S_j  \right )
\left( A^l+B^l-2C^l \right ) , 
\label{eq:ha2}
\end{eqnarray}
where $\vec S_i$ is the spin operator with magnitude $S=1$. 
$J_{^4A_2}(=t^2/(U'-I) \equiv J)$ is the exchange interaction 
with the on-site inter-orbital Coulomb interaction $U'$,  
the exchange interaction $I$ and the transfer intensity $t$ between NN V ions. 
The other exchange interactions are expressed by 
$J$ and $r_J=I/U'$. 
The orbital sector in ${\cal H}_{^4J_2}$ involving $A^l$, $B^l$, and $C^l$,  
where $l(=x,y,z)$ indicates a direction of a bond, 
is expressed by the local orbital operators at sites $i$ and $j$.  
These describe the diagonal and off-diagonal 
components of the electric-quadrupole moments, 
$O_{i E \gamma} \ (\gamma=u,v)$ and 
$O_{i T_2 \gamma}\ (\gamma=x,y,z)$, respectively, 
and the orbital angular moments   
$O_{i T_1 \gamma}\ (\gamma=x,y,z)$. 
There is no continuous symmetry 
in the orbital sector of ${\cal H}_J$. 
The $t$ term is given as    
\begin{equation}
{\cal H}_t=\sum_{\langle ij \rangle, \gamma, \sigma} 
\left ( t_{ij}^\gamma 
c_{i \gamma \sigma}^\dagger 
c_{j \gamma \sigma}^{}+H.c. \right ) .
\label{eq:ht}
\end{equation}
The operator $c_{i \gamma \sigma}$ 
annihilates a $t_{2g}$ electron at site $i$ 
with orbital $\gamma=(yx,zx,xy)$ 
and spin $\sigma=(\uparrow, \downarrow)$, 
and is defined in the Hilbert space where the $(t_{2g})^3$ 
configuration is excluded at each site. 
In the ideal perovskite crystal,
the hopping between the different orbitals is prohibited, 
and 
one of $t_{ij}^\gamma$'s  
is zero and other two are equal, e.g. 
$t^{xy}_{i, i+\hat z}=0$ and $t^{yz}_{i, i+\hat z}=t^{zx}_{i,  i+\hat z} \equiv t$. 
In addition to the $t$ and $J$ terms, 
roles of the Jahn-Teller (JT) interaction described by 
${\cal H}_{JT}=g \sum_{i \gamma} Q_{i E \gamma} O_{i E \gamma}$  
are stressed in the insulating $R$VO$_3$ \cite{motome}. 
$Q_{i E \gamma}$'s $(\gamma=u,v)$ are the normal modes of an O$_6$ 
octahedron with symmetry $E_g \gamma$ and are treated as classical valuables. 
The detail expression of the above Hamiltonian 
and studies of the orbital order/excitation  
are presented in Refs.~\cite{ishihara04} and \cite{ishihara02}. 
Similar spin-orbital models for insulating 
$R$VO$_3$ are derived by other authors \cite{khaliullin01,horsch,motome,shen,kugel,harris}. 

Dynamics of holes doped into the spin-orbital ordered vanadates 
are formulated by SCBA 
which has been applied to HTSC  
\cite{kane,khaliullin} 
and the colossal magnetoresistive (CMR) manganites \cite{brink,yin,bala}. 
The electron operator is expressed in the slave-fermion representation 
in the large limit of $S$. 
We assume that the orbital excitation  
between the $d_{yz}$ and $d_{zx}$ orbitals 
couples effectively to the doped holes, since 
(i) the excitation between $d_{xy}$ and $d_{yz}/d_{zx}$ 
dose not propagate coherently (local modes) \cite{ishihara04},  
and (ii) the $d_{xy}$ band is separated from 
the $d_{yz}/d_{zx}$ ones, i.e. 
the excitation energy between $d_{xy}$ and $d_{yz}/d_{zx}$ 
is higher \cite{sawada}. 
The electron operator is rewritten as,  
\begin{equation}
c_{i \gamma \sigma}=
\left [
\delta_{\gamma=(xy)} {\tilde f}_i^{} {\tilde s}_{i \sigma}^{\dagger}
+\delta_{\gamma=(zx,yz)} f_{i}^{} s_{i \sigma}^{ \dagger} t_{i \gamma}^\dagger 
\right ]
P_S . 
\label{eq:dtilde}
\end{equation}
We introduce two fermionic operators $\{ {\tilde f}_i, f_i \}$ for charge, 
two bosonic operators $\{ {\tilde s}_{i \sigma}, s_{i \sigma} \}$ 
for spin, and 
bosonic one $\{ t_{i \gamma} \}$ for orbital 
with the local constraints 
${\tilde f}^\dagger_i {\tilde f}_i^{}
+\sum_{\sigma} {\tilde s}_{i \sigma}^\dagger {\tilde s}_{i \sigma} =1$, 
${f}_i^\dagger {f}_i^{}+\sum_{\sigma} {s}_{i \sigma}^\dagger {s}_{i \sigma} =1$ 
and 
$\sum_\sigma s_{i \sigma}^\dagger s_{i \sigma}=
\sum_{\gamma=(yz,zx)} t_{i \gamma}^\dagger t_{i \gamma}$. 
The operator $P_S$ projects onto the $^3T_1$ high-spin state 
in the $(t_{2g})^2$ configuration. 
Then, ${\cal H}_t$ is decoupled into the $d_{yz}/d_{zx}$ orbital part and 
the $d_{xy}$ one as 
${\cal H}_t={\cal H}_{yz/zx}+{\cal H}_{xy}$ 
with ${\cal H}_{yz/zx}={\cal H}^{(c)}+{\cal H}^{(ab)}$. 
The main term is ${\cal H}^{(c)}$ describing a  
motion of the $d_{yz}/d_{zx}$ hole 
along the $c$ axis of our present interest. 
In the large limit of $S$, we remain the 
down (up) boson operators 
in the up (down) spin/orbital sublattice. 
${\cal H}^{(c)}$ has the form 
\begin{eqnarray}
{\cal H}^{(c)}=
\frac{4t}{\sqrt{N}} 
\sum_{\vec k, \vec k'} g(k_z)
f_{\vec k}^{} f_{\vec k'}^\dagger
u_{\vec k' -\vec k}^{}+H.c. , 
\end{eqnarray}
where $u_{\vec k}=s_{\vec k}$ for the (SG/OC) phase and  
%\begin{eqnarray}
%{\cal H}^{(c)}_{\rm (SC/OG)}=
%\frac{4t}{\sqrt{N}} 
%\sum_{\vec k, \vec k'} g(k_z)
%f_{\vec k}^{} f_{\vec k'}^\dagger
%t_{\vec k'-\vec k }^{}+H.c. , 
%\end{eqnarray}
$u_{\vec k}=t_{\vec k}$ for (SC/OG) with $g(k_l)=\cos ak_l$. 
Explicitly, the mobile hole 
along the $c$ axis is coupled to magnon (orbiton) in 
the (SG/OC) ((SC/OG)) phase. 
${\cal H}^{(ab)}$ and ${\cal H}_{xy}$,  
are common in the two phases. 
${\cal H}^{(ab)}$ is for the $d_{yz}/d_{zx}$ hole in the $ab$ plane,  
and fomrs 
$f_{\vec k}^{} f_{\vec k'}^\dagger
s_{\vec q}^{\dagger} t_{\vec k'-\vec k+\vec q}^\dagger$.
The simultaneous magnon-orbiton emissions/absorptions 
is expected to suppress the QP weight and dispersion. 
${\cal H}_{xy}$ is for the $d_{xy}$ hole in the $ab$ plane, 
%\begin{eqnarray}
$
{\cal H}_{xy}=\frac{4t}{\sqrt{N}} 
\sum_{\vec k, \vec k'} 
\{ g(k_x)+g(k_y) \} 
{\tilde f}_{\vec k}^{} {\tilde f}_{\vec k'}^\dagger 
{\tilde s}_{\vec k-\vec k' }  +H.c. . 
$
%\end{eqnarray}
Similar physical consequences are only to be expected from 
this term to those in a mobile $d_{x^2-y^2}$ hole 
in the 2D quantum AFM 
studied in HTSC \cite{kane}. 
Here we focus on ${\cal H}^{(c)}$ 
characterizing different hole dynamics 
in the (SG/OC) and (SC/OG) phases. 
In the same scheme, 
${\cal H}_J$ is formulated 
by the spin wave expansion 
\cite{ishihara04}. 
In the magnon (orbiton) part of ${\cal H}_J$
written by $s_{i \sigma}$ ($t_{i \gamma}$), 
the orbital (spin) operators are replaced by 
the static correlation functions 
$\langle O_{i \Gamma \gamma}O_{j \Gamma' \gamma'}\rangle$ 
($\langle \vec S_i \cdot \vec S_j \rangle$), 
and the simultaneous spin-orbital excitations are neglected, 
same as those in ${\cal H}_t$ mentioned above. 
The lattice effects are limited to be static ones 
given by ${\cal H}_{JT}$, 
although the coupling with phonon may modify the QP band width and 
the orbiton dispersion around the crossing point with phonon 
in the small $J$ limit. 

\begin{figure}
\epsfxsize=0.75\columnwidth
\centerline{\epsffile{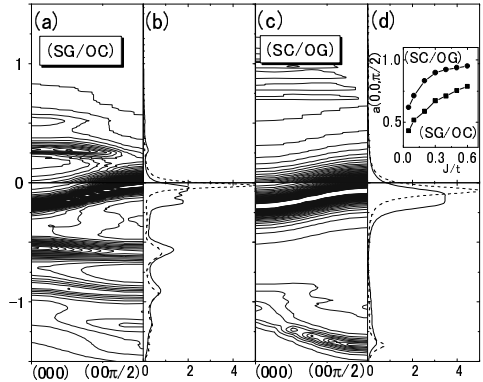}}
\caption{Contour plot of spectral function  
and DOS of electron 
in (SG/OC) ((a) and (b)), and those 
(SC/OG) ((c) and (d)) for $x=0.1$. 
Broken lines in (b) and (d) are for 
$A(\vec k=(0,0,\frac{\pi}{2a}),\omega)$ 
multiplied by 1/5.
Parameters are $J/t=0.2$, $r_J=0.125$ and $gQ_{E}/J=1.25$. 
The origin of the vertical axis indicates the Fermi energy.
The inset shows the $J$ dependence of the QP weight at 
$\vec k=(0,0,\frac{\pi}{2a})$.}
\label{fig1}
\end{figure}
The electron spectral function $A(\vec q, \omega)$ is 
calculated from the fermion propagator $G(\vec k, \omega)$
where the boson operators are replaced by 
the saddle-point solutions.  
%with 
%$G(\vec k, \omega)=-i
%\int dt e^{i \omega t} \langle T  f_{\vec k }^{}(t) f_{\vec k }^{\dagger}(0) \rangle$. 
The self-energy for $G(\vec k, \omega)$ is given by 
\begin{eqnarray}
\Sigma(\vec k, \omega)&=&\frac{i}{2\pi N} \int d \nu \sum_{\vec q}
G(\vec k-\vec q, \omega-\nu) 
\nonumber \\
& \times & 
|{\widetilde g}_u(\vec k, \vec q)|^2
D_{u0}(\vec q, \nu)  , 
\label{eq:sigma}
\end{eqnarray}
where $u=s$ for (SG/OC) and $u=t$ for (SC/OG). 
The bare magnon/orbiton propagators $D_{u0}(\vec q, \nu)$ 
are used in Eq.~(\ref{eq:sigma}) instead of the full propagators 
$D_{u}(\vec q, \nu)$. 
${\widetilde g}_u(\vec k, \vec q)$ is the coupling constant 
including the coefficients in the Bogoliubov transformation. 
%$U_u(\vec q)$ indicates the coefficient in the Bogoliubov transformation   
%and ${\widetilde g}_u(\vec k, \vec q)=g(k_z-q_z)$ for $ u_{ \vec q} $ 
%and $g(k_z)$ for $u_{m -\vec q}^\dagger$.  
%
$A(\vec k, \omega)$ along 
$(0,0,0)-(0,0,\frac{\pi}{2a})$, with $a$ being the 
cubic perovskite unit cell, 
and the density of states (DOS) $N(\omega)$ are obtained in the  
self-consistent calculation (Fig.~1). 
A reasonable parameter set as  
$J/t=0.2$, $r_J=0.125$ and  $gQ_{E}/J=1 \sim 1.25$ with 
$J$ being around 30meV 
is chosen based on the previous evaluations \cite{motome,ishihara04}. 
In both the two phases, 
$A(\vec k, \omega)$ shows the broad incoherent continuum 
and the well-separated QP peak near the Fermi level (FL). 
The dispersive QP peak has the band width $D$ of the order of $J$ 
and has the highest energy at the momentum $(0,0,\frac{\pi}{2a})$.  
The clear differences 
between the two phases are seen 
in the QP spectral weight $a(\vec k)$ and $D$;  
in the (SC/OG) phase, a large portion of the spectra are  
concentrated on the QP peak with the larger width $D$,  
and $a(\vec k)$ is rapidly saturated with increasing 
$J$ (the inset of Fig.~1). 
This is in contrast to the case of (SG/OC)  
where the spectra are dominated by the incoherent components. 
These are ascribed to the larger excitation energy of orbiton with 
the excitation gap about $2J \sim 3J$ 
(as shown later in Fig.~3(b)). 
We have checked and confirmed these features 
of $a(\vec k)$ and $D$ by the  
variational-type calculations \cite{hatakeyama}. 
It is worth noting that 
the hole dynamics associated with the $t_{2g}$ 
orbitons in (SC/OG) is in 
qualitatively contrast to 
that with the $e_g$ orbital for the CMR manganites  
\cite{koshibae,brink};  
$A(\vec k, \omega)$ has a large weight around 
the free $e_g$ bands rather than the QP part. 
This is because of the non-zero hopping matrices between 
the NN different $e_g$ orbitals unlike the $t_{2g}$ ones. 

\begin{figure}
\epsfxsize=0.7\columnwidth
\centerline{\epsffile{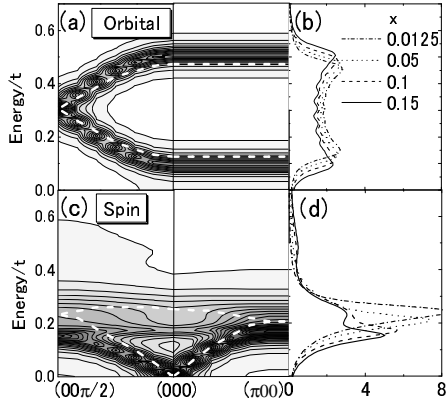}}
\caption{Contour plots of the spectral functions at $x=0.1$ 
and the doping dependence of DOS for orbiton ((a) and (b)) and 
those for magnon ((c) and (d)) in (SG/OC).
White broken lines in (a) and (c) indicate dispersions at $x=0$. 
The lower branch of the orbiton in (a) 
corresponds to that along $(0,0,\frac{\pi}{2a})-(0,0,\frac{\pi}{a})-(\frac{\pi}{a},0,\frac{\pi}{a})$ 
in the Brillouine zone for the cubic unit cell. 
Parameters are the same with those in Fig.~1. 
}
\label{fig2}
\end{figure}
Now we turn to the spin and orbital sectors. 
The renormalized magnon/orbiton propagators are given 
by utilizing the calculated  $G(\vec k, \omega)$.  
The bubble-type diagram is the lowest order of the self-energy, 
\begin{eqnarray}
\Pi_u(\vec q, \omega)&=&
\frac{-i}{2 \pi N} \int d \nu \sum_{\vec k}
|{\widetilde g}_u^t(\vec k, \vec q)|^2 
\nonumber \\
&\times& 
G(\vec k+\vec q, \omega+\nu)  G(\vec k, \nu)   . 
\end{eqnarray}
The spectral functions $S(\vec q, \omega)=(-1/\pi){\rm Im}D_s(\vec q, \omega)$ 
and $T(\vec q, \omega)=(-1/\pi){\rm Im}D_t(\vec q, \omega)$, 
and DOS in (SG/OC) are shown in Fig.~2. 
In the undoped case (white broken lines), 
the calculated $S(\vec q, \omega)$ well reproduces 
the experimentally observed spin wave dispersion in YVO$_3$ \cite{ulrich}. 
The flat dispersion of $T(\vec q, \omega)$ in the $ab$ plane 
originates from no coherent propagation of the 
$d_{yz}/d_{zx}$ orbital excitations 
in this plane \cite{ishihara04}. 
The excitation gap in $T(\vec q, \omega)$ at $x=0$ 
comes from the discontinuous symmetry of 
the orbital sector in ${\cal H}_J$ and the JT interaction, 
and is much smaller than that in (SC/OG). 
Namely, the fragile character 
is inherent in the C-type orbital order. 
As expected from distortions of the 
staggered spin/orbital array by mobile holes,  
softening and broadening are observed for  
$S(\vec q, \omega)$ in (SG/OC) (Figs.~2(c) and 2(d)) 
and for $T(\vec q, \omega)$ in (SC/OG) (not shown). 
Unexpectedly,  
the band width of $T(\vec q, \omega)$ in (SG/OC) 
becomes broader with doping (Figs.~2(a) and 2(b)), 
in spite that the orbital distortion due to the hole motion 
is not expected along the $c$ axis. 
The minimum energy $\omega_{OW}^{{\rm Min}}$  
of $T(\vec q, \omega)$ in this phase 
is located at the point $\Gamma$ in the lowest branch 
which corresponds to 
the point $(0,0,\frac{\pi}{a})$ 
in the original Brillouine zone for the cubic perovskite unit cell. 
$\omega_{OW}^{{\rm Min}}$  
decreases with doping and finally touches the zero energy at small number of $x$. 
That is, the (SG/OC) phase is unstable against the small doping  
and is changed into the G-type orbital phase. 
\begin{figure}
\epsfxsize=0.8\columnwidth
\centerline{\epsffile{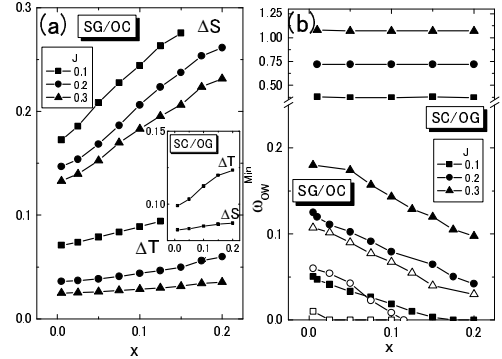}}
\caption{(a) Doping dependences of 
reductions of the spin and orbital order parameters, 
$\Delta S$ and $\Delta T$, in (SG/OC). 
The inset shows those in (SC/OG) with $J/t=0.2$. 
(b) Doping dependence of the minimum  
orbiton energy $\omega_{OW}^{{\rm Min}}$. 
The filled and open symbols in (b) are for $gQ_E/J=$1.25 and 1, respectively. }
\label{fig3}
\end{figure}
This orbiton softening in (SG/OC) originates from 
reduction of the staggered spin order parameter, 
termed $\Delta S(>0)$ (see Fig.~3). 
The remarkable change by doping 
is only observed in $\Delta S$ of the (SG/OC) phase.  
However, this reduction itself is not sufficient for 
instability of the AFM ordering, i.e. $\Delta S < S(=1)$. 
Instead, 
the effective exchange interaction for 
the staggered (uniform) orbital alignment in ${\cal H}_J$, 
given as 
$J_{AF}^{orb}=J_{^4A_2}(2+\langle \vec S_i \cdot \vec S_j \rangle)$ 
($J_{F}^{orb}=J_{\Gamma}(1-\langle \vec S_i \cdot \vec S_j \rangle)$ 
($\Gamma=$$^2E$, $^2T_1$, $^2T_2$)), 
rapidly increases (decreases) with increasing $\Delta S$. 
Finally, the orbiton energy becomes zero 
at the critical $x$ 
which depends on the exchange and JT interaction parameters. 
In short, the instability of the (SG/OC) phase with doping, 
observed in Y$_{1-x}$Ca$_x$VO$_3$ \cite{miyasaka_un}, 
is caused by the orbiton softening due to 
increasing of $\Delta S$ through the spin-orbital coupling. 

The qualitatively different hole dynamics 
in the two phases can be probed 
by the optical experiments. 
The $z$-component of the 
optical conductivity spectra $\sigma_{zz}(\omega)$ 
is formulated 
in the slave-fermion scheme with the $1/S$ expansion. 
The lowest order diagram of $\sigma_{zz}(\omega)$ is given by 
$\sigma_{zz}(\omega)=-\frac{(9te)^2}{c^2} {\rm Im} K(\omega)$
with 
\begin{eqnarray}
K(\omega)&=&\frac{2}{(2\pi N)^2 \omega} 
\int d \omega_1 d \omega_2 \sum_{\vec k_1 , \vec k_2}
|{\widetilde g}_u(\vec k_1,\vec k_2)|^2 G(\vec k_1, \omega_1)
\nonumber \\
&\times &
D_{u 0}(\vec k_2-\vec k_1, \omega-\omega_1+\omega_2)
 G(\vec k_2, \omega_2) , 
\end{eqnarray} 
for $u=s$ ($u=t$) in the (SG/OC) ((SC/OG)) phase, 
indicating the particle-hole 
pair creation associated with magnon/orbiton. 
Here we introduce the results of the regular part of $\sigma_{zz}(\omega)$ 
at finite frequency, rather than the Drude part (Fig.~4). 
With increasing holes, 
the sharp peak structure growing up around $\omega =J \sim 2J$ 
in the (SC/OG) phase, reflecting the large QP weight 
in $A(\vec k, \omega)$ (Fig.~1), 
in contrast to the broad incoherent spectra in (SG/OC).  
A shoulder structure is clearly seen 
in the integrated spectral weight 
$M(\omega)=\frac{c^2}{(9te)^2}\int_0^\omega \sigma_{zz}(\omega') d \omega'$ 
(the inset of Fig.~4) 
which would be an evidence characterizing the hole dynamics 
in the orbital ordered insulators.

In summary, we present a theory of doped vanadium perovskites 
with spin and $t_{2g}$ orbital orders. 
The doped (SG/OC) and (SC/OG) phases are characterized 
by the hole motion along the $c$ axis scattered by  
magnon and orbiton, respectively. 
The QP dynamics associated with the $t_{2g}$ orbital excitation 
show remarkable difference from that with magnon as well as 
the $e_g$ orbital excitation. 
Instability of the (SG/OC) phase into (SC/OG) 
observed in Y$_{1-x}$Ca$_x$VO$_3$ is attributed to the 
orbiton softening through a reduction of the staggered spin order parameter. 
The angular resolved photoemission spectroscopy and optical experiments for 
the QP, as well as the inelastic neutron scattering for 
magnon and orbiton \cite{ishihara04}, are required to test directly the present theoretical results. 
\begin{figure}
\epsfxsize=0.75\columnwidth
\centerline{\epsffile{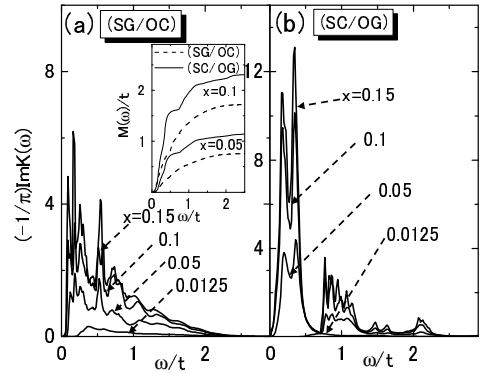}}
\caption{Regular parts of the optical conductivity spectra 
for (SG/OC) (a), and those for (SC/OG) (b).
The inset shows the integrated spectral weights for the (SG/OC) phase 
(broken lines) and those for the (SC/OG) ones (bold lines).  
Parameter values are the same with those in Fig.~1.
}
\label{fig4}
\end{figure}

Author would like to thank T.~Hatakeyama and S.~Maekawa 
for their valuable discussions, 
and S.~Miyasaka and Y.~Tokura   
for providing their unpublished data. 
This work was supported by KAKENHI from MEXT, and 
KURATA foundation. 
Part of the numerical calculation has been performed by 
the supercomputers in IMR, Tohoku Univ., 
and ISSP, Univ. of Tokyo. 


\begin{references}
\bibitem{tokura}
M.~Imada, {\it et al.},  
Rev. Mod. Phys. {\bf 70}, 1039 (1998).  
%\bibitem{tokura00}
%Y.~Tokura and N.~Nagaosa, 
%Science {\bf 288}, 462 (2000). 
\bibitem{book}
S.~Maekawa, {\it et al.}, 
{\it Physics of Transition Metal Oxides}, 
(Springer Verlag, Berlin, 2004) and references therein. 
%\bibitem{mott}
%N.~F.~Mott, 
%{\it Metal-Insulator Transitions}, 
%(Talyor $\&$ Francis, London, 1990). 
%
\bibitem{ren}
Y.~Ren {\it et al.}, 
Nature {\bf 396}, 441 (1998). 
\bibitem{miyasaka00}
S.~Miyasaka {\it et al.}, 
%T.~Okuda, and Y.~Tokura, 
Phys. Rev. Lett. {\bf 85}, 5388 (2000). 
%\bibitem{noguchi}
%N.~Noguchi {\it et al.}, 
%Phys. Rev. B {\bf 62}, R9271 (2000).
%\bibitem{blake}
%G.~R.~Blake {\it et al.}, 
%Phys. Rev. Lett. {\bf 87}, 245501 (2001). 
%\bibitem{miyasaka02}
%S.~Miyasaka {\it et al.}, 
%T.~Okuda, and Y.~Tokura, 
%Jour. Phys. Soc. Jpn. {\bf 71}, 2086 (2002). 
\bibitem{miyasaka03}
S.~Miyasaka {\it et al.}, 
%Y.~Okimoto, M.~Iwasa, and Y.~Tokura, 
Phys. Rev. B {\bf 68}, 100406 (2003).
%
\bibitem{sawada}
H.~Sawada {\it et al.},
Phys. Rev. B {\bf 53}, 12742 (1996). 
%\bibitem{mizokawa}
%T.~Mizokawa, {\it et al.}, 
%and A.~Fujimori, 
%Phys. Rev. B {\bf 54}, 5368 (1996). 
%
\bibitem{khaliullin01}
G.~Khaliullin {\it et al.}, 
%P.~Horsch, and A.~M.~Oles
Phys. Rev. Lett. {\bf 86}, 3879 (2001). 
\bibitem{horsch}
P.~Horsch {\it et al.}
% G.~Khaliullin, and A.~M.~Oles, 
Phys. Rev. Lett. {\bf 91}, 257203 (2003). 
\bibitem{motome}
Y.~Motome {\it et al.}
% H.~Seo, Z.~Fang, and N.~Nagaosa, 
Phys. Rev. Lett. {\bf 90}, 146602 (2003). 
\bibitem{shen}
S.-Q.~Shen {\it et al.}
%X.~C.~Xie, and F.~C.~Zhang, 
Phys. Rev. Lett. {\bf 88}, 027201 (2002). 
\bibitem{ulrich}
C.~Ulrich {\it et al.},
Phys. Rev. Lett. {\bf 91}, 257202 (2003). 
\bibitem{ishihara04}
S.~Ishihara, Phys. Rev. B {\bf 69}, 075118 (2004).
\bibitem{saitoh}
E.~Saitoh {\it et al.}, Nature, {\bf 410}, 180 (2001). 
%
\bibitem{miyasaka_un}
S.~Miyasaka and Y.~Tokura,  
(unpublished). 
In YVO$_3$, the high temperature (SG/OC) phase is observed between 
75K and 120K, and the low temperature (SC/OG) one is below 75K. 
At around $x=0.02$ in Y$_{1-x}$Ca$_{x}$VO$_3$,  
the (SC/OG) phase is changed into (SG/OC) 
which survives, at least, above $x=0.4$. 
%The (SG/OC) phase observed in the LaVO$_3$ below 145K
%is changed into the 
%AF metallic phase at $x=0.175$ in La$_{1-x}$Sr$_{x}$VO$_3$. 
%
%\bibitem{abc}
%The orbital parts of Eq.~(\ref{eq:ha2}) are explicitly given by 
%$A^l=-4O_{iEv}^l O_{j Ev}^l$, 
%$B^l=V_{i}^lW_{j}^l+W_{i}^lV_{j}^l$, and 
%$C^l=2(O_{iT_2 l} O_{jT_2 l}+O_{i T_1 l} O_{j T_1 l})$ 
%with 
%$W^l_i=(2/3)-\sqrt{2/3}O_{iE_u}^l$ and $V_i^l=(1/3)-\sqrt{2/3}O_{iE_u}^l$. 
%$O_{iEu(v)}^l$ are 
%$O_{iEu}^l=\cos(2\pi m_l/3)O_{iEu}+\sin(2\pi m_l/3)O_{iEv}$
%and 
%$O_{iEv}^l=-\sin(2\pi m_l/3)O_{iEu}+\cos(2\pi m_l/3)O_{iEv}$ 
%with $m_l=(1,2,3)$ for $l=(x,y,z)$. 
%
\bibitem{ishihara02}
S.~Ishihara {\it et al.},
%T.~Hatakeyama, and S.~Maekawa, 
Phys. Rev. B {\bf 65}, 064442 (2002). 
%
\bibitem{kugel}
K.~I.~Kugel {\it et al.}, 
%and D.~I.~Khomskii, 
%Fiz. Tverd. Tela (Leningrad) {\bf 17}, 454 (1975) 
Sov. Phys. Solid State {\bf 17}, 285 (1975). 
\bibitem{harris}
A.~B.~Harris {\it et al.}, 
Phys. Rev. Lett. {\bf 91}, 087206 (2003). 
%\bibitem{schmitt}
%S.~Schmitt-Rink, {\it et al.}, 
%Phys. Rev. Lett. {\bf 60}, 2793 (1988). 
\bibitem{kane}
C.~L.~Kane {\it et al.}, 
Phys. Rev. B {\bf 39}, 6880 (1989). 
%\bibitem{ramsak}
%A.~Ramsak, and P.~Prelovsek, 
%Phys. Rev. B {\bf 42}, 10415 (1990). 
%\bibitem{martinez}
%G.~Mart$\rm \acute{ i}$nez and P.~Horsch, 
%Phys. Rev. B {\bf 44}, 317 (1991). 
%\bibitem{igarashi}
%J.~Igarashi, and P.~Fulde, 
%Phys. Rev. B {\bf 48}, 12713 (1993).  
%{\it ibid}, {\bf 45}, 12537 (1992). 
\bibitem{khaliullin}
G.~Khaliullin {\it et al.}, 
%and P.~Horsch, 
Phys. Rev. B {\bf 47}, 463 (1993). 
%
\bibitem{brink}
J.~van~den~Brink {\it et al.}, 
Phys. Rev. Lett. {\bf 85}, 5174 (2000). 
\bibitem{yin}
W.~G.~Yin {\it et al.}, 
Phys. Rev. Lett. {\bf 87}, 047204 (2001). 
\bibitem{bala}
J.~Bala {\it et al.}, 
Phys. Rev. Lett. {\bf 87}, 067204 (2001). 
%\bibitem{allen}
%P.~B.~Allen, and V.~Perebeinos, 
%Phys. Rev. B {\bf 60}, 10747 (1999). 
%\bibitem{eq1}
%Sets of the operators,  
%$\{ f_{m \vec k} \}$ and $\{ u_{m \vec q}, u_{m -\vec q}^\dagger \}$ with $(m=1 \sim 4)$,  
%for the four V ions in a unit cell 
%are treated by $f_{\vec k}$ and $u_{\vec q}$, respectively, 
%and the indices $m$ is omitted. 
\bibitem{hatakeyama}
A part of the results is presented in 
S.~Ishihara {\it et al.}, 
%and T.~Hatakeyama, 
Jour. Mag. Mag. Mat. {\bf 272-276}, 412 (2004). 
%\bibitem{inoue}
%J.~Inoue, and S.~Maekawa, 
%Jour. Phys. Soc. Jpn. {\bf 59}, 2110 (1990). 
\bibitem{koshibae}
W.~Koshibae {\it et al.}, 
%and S.~Maekawa, 
Physica C {\bf 317-318}, 205 (1999).
%
\end{references}
\end{document}